\documentclass[conference]{IEEEtran}

\usepackage[utf8]{inputenc}
\usepackage[T1]{fontenc}
\usepackage[australian]{babel}
\usepackage{color,xcolor}
\usepackage{listings}

\usepackage{amsmath}
\usepackage{marginnote}
\usepackage{colortbl}
\usepackage{multirow}
\usepackage[disable]{todonotes}

\PassOptionsToPackage{hyphens}{url}
\usepackage{enumitem}
\usepackage{subfigure}
\usepackage{datetime}
\usepackage{balance}
\usepackage{algorithm}
\usepackage{algpseudocode}

\usepackage{epstopdf}
\usepackage{booktabs}
\usepackage{multicol}
\usepackage{hyperref}
\hypersetup{
    colorlinks = true,
    linkcolor = {blue},
    linkbordercolor = {blue},
    anchorcolor = {blue},
    citecolor = {blue},
}

\graphicspath{ {./} }

\begin{document}

\title{Double-Spending Risk Quantification in Private, Consortium and Public Ethereum Blockchains}

\author{
	\IEEEauthorblockN{Parinya Ekparinya}
	\IEEEauthorblockA{
				University of Sydney, Australia\\
		Email: pekp6601@uni.sydney.edu.au}
	\and
	\IEEEauthorblockN{Vincent Gramoli}
	\IEEEauthorblockA{
				University of Sydney, Australia\\
				Data61-CSIRO, Sydney, Australia \\
		Email: vincent.gramoli@sydney.edu.au}
	\and
	\IEEEauthorblockN{Guillaume Jourjon}
	\IEEEauthorblockA{
				Data61-CSIRO, Sydney, Australia\\
		Email: guillaume.jourjon@data61.csiro.au}
}



\maketitle
\begin{abstract}
Recently, several works conjectured the vulnerabilities of mainstream blockchains under several network attacks.
All these attacks translate into showing that the assumptions of these blockchains can be violated in theory or under simulation at best.
Unfortunately, previous results typically omit both the nature of the network under which the blockchain 
code runs and whether blockchains are private, consortium or public.

In this paper, we study the public Ethereum blockchain as well as a consortium and private blockchains and quantify the feasibility of man-in-the-middle and double spending attacks against them.
To this end, we list important properties of the Ethereum public blockchain topology, we deploy VMs with constrained CPU quantum to mimic the top-10 mining pools of Ethereum and we develop full-fledged attacks, that first partition the network through BGP hijacking or ARP spoofing before issuing a Balance Attack to steal coins.
%
Our results demonstrate that attacking Ethereum is remarkably devastating in a consortium or private context as the adversary can multiply her digital assets
by $200,000\times$ in 10 hours through BGP hijacking whereas it would be almost impossible in a public context.
\end{abstract}

\section{Introduction}
\label{sec:intro}


Blockchains, like Bitcoin~\cite{Nak08} or Ethereum~\cite{Woo15}, implement cryptocurrencies through a distributed system that relies heavily on network communications.
%
As the price of these 
cryptocurrencies have skyrocketed in the recent years, we are facing a growing amount of network attacks
to steal their corresponding coins.
These attacks are not new as in 2014,
an attacker acted as a man-in-the-middle by hijacking BGP routes in one of the Canadian autonomous systems (ASes)  
to steal US\$ 83,000. 
Few days ago, another BGP hijacking attack against an Ethereum wallet allowed an attacker to steal Ethereum coins or \emph{ethers}.\footnote{\url{https://www.theregister.co.uk/2018/04/24/myetherwallet_dns_hijack}.}
Although this attack is simpler as it targeted a single server, it illustrates the proliferation of network attacks against blockchain.

In fact, communication delays can typically be exploited to steal assets from blockchains.
As transactions and blocks that update the state of the blockchain are sent through the network, 
distant nodes can observe conflicting transactions while these transaction blocks are being propagated to all nodes.
Consider for example that  a merchant observing a transaction $t$ decides to
transfer a physical asset in the real world before receiving the up-to-date state of the blockchain. 
If this transaction $t$ gets finally discarded due to an existing conflict, then the  merchant loses its asset and the attacker can re-spend the coins he spent in $t$, hence the name \emph{double spending}.
Double spending remains a critical vulnerability of most blockchains that rely on proof-of-work and 
resulted again in asset losses in the last few weeks.\footnote{\url{https://www.newsbtc.com/2018/04/05/cryptocurrency-verge-has-been-hit-with-a-51-attack-loses-250000-tokens/}.}

Several research results confirmed that network delays affect ``in theory'' the security of blockchain systems.
In particular, it is well known that delaying network messages can impact Bitcoin~\cite{DW13,PSS16,SZ15,GKK16,NKMS16} but only few results tried to assess the vulnerability of Ethereum~\cite{NG16,NG17}.
Some attacks rely on the simple idea, called solo-mining or selfish mining, of delaying the propagation of blocks already mined~\cite{Fin11,Vec11}.
Other attacks require to delay the messages between one node and the rest of the network~\cite{HKZG15,NG16}.
Finally, some attacks involve partitioning the network by using man-in-the-middle attacks~\cite{AZV17,NG17}.
To our knowledge, however, there is no full-fledged attack that combines \textit{(i)}~a real network attack to delay the messages and then  \textit{(ii)}~a double spending attack leveraging these delays to steal from the blockchain. Current approaches typically focus on either one of these two aspects.


%
By focusing only on the network attack feasibility in blockchain, one can ignore the feasibility of generally stealing digital assets in realistic settings.
First, such results would extrapolate the applicability of theoretical network attacks to the public context and 
it remains unclear to which extent a real attack can succeed in a public or consortium environment.
Second, the results that demonstrate the feasibility of a network attack experimentally
%
%
would ignore the
pseudo random process used at the heart of proof-of-work blockchains that could presumably translate these attacks into successful asset losses.
While network attacks may implicitly contribute to the risks of asset losses, 
the extent to which this is possible depending on the environment remains unclear.

In this paper, we quantify empirically the risks for an adversary to steal coins in Ethereum by executing \textit{(i)}~a man-in-the-middle attack followed by a \textit{(ii)}~double spending attack. 
To this end, we gathered connectivity and mining power information of the Ethereum public blockchain, and setup an Ethereum sandboxed testnet to mimick public, consortium and private environments. 
%
Our testnet comprises BGP routers and VMs configured with OpenStack. We reflected the mining power of the top-10 Ethereum mining pools by restricting the CPU quantum of each VM with linux \texttt{cgroups}. We then performed 
BGP hijacking and ARP spoofing to try partitioning the networks before running a Balance Attack~\cite{NG17} and measuring empirically the risks of double spending.

Our first conclusion is that stealing assets against public blockchains is very hard due to the nature of the network topology. 
Our second conclusion is that attacks against consortium and private blockchains are surprisingly easy and lead to dramatic losses.
Finally, we quantify the asset gain of an adversary that executes the attack continuously for a period of time and we discuss the countermeasures to lower the risks of man-in-the-middle attacks against Ethereum.
Within roughly 10 hours, the adversary could gain as much as 200,000 folds of their initial funding.

The rest of the paper is organized as follows.
We present the background in Section~\ref{sec:background}.
We discuss in Section~\ref{sec:problem} why network attacks against public Ethereum are not trivial.
We describe the experimental results of double spending with BGP-Hijacking in a consortium deployment in Section~\ref{sec:consortium}.
We describe the experimental results of double spending through ARP-spoofing in a private context in Section~\ref{sec:private}.
We discuss the impact and also countermeasures of the attacks in Section~\ref{sec:discussion}.
In Section~ \ref{sec:related}, we present related work from past research.
Finally, we conclude our paper in Section~ \ref{sec:conclusion}.

\section{Background and Motivations}
\label{sec:background}

A blockchain system offers a tamper-proof ledger~\cite{underwood_blockchain_2016} distributed on a collection of communicating nodes, all sharing the same initial block of information, the \emph{genesis} block.
In order to add information to the blockchain, a node includes information in a block with a pointer to its parent block, this creates a chain of blocks, hence called \emph{blockchain}.
To create a block, a node usually needs to solve a crypto-puzzle and provides the solution as a proof of its work to get a reward, this process is called \emph{mining}~\cite{Nak08}. 
The difficulty of the crypto puzzle is adjusted based on the total computational power or \emph{mining power} of the blockchain network.
Each correctly behaving miner needs to adhere to the same protocol for creating and also validating new blocks.
Upon successfully mining a block, a miner broadcasts it for validation. 

Unfortunately, network delays impact the time to hear about new blocks and during that time miners may append multiple blocks pointing to the same parent blocks, a situation called a \emph{fork}.
Such forks
may lead to different nodes accepting conflicting information. 
If these forks persist, they may lead to \emph{double spending}~\cite{Fin11,Ros12}, where an adversary spends the same coins in transactions located in two branches of the forked blockchain.
Intuitively, the longer the communication delay, the longer the fork will remain undetected and persist.
%
To resolve forks, blockchain platforms implement a blockchain consensus protocols to select a single branch as a canonical chain.
One way to resolve the issue is to simply choose the longest branch whenever there are forks, leaving blocks that are not part of the main chain as stale blocks~\cite{Nak08}.
Other variants exist, for example the {\sc Ghost} protocol~\cite{SZ15} that initially influenced the Ethereum consensus protocol~\cite{Woo15} selects a canonical branch by considering total weight of the subtrees including stale blocks.

\subsection{Ethereum} Ethereum~\cite{Woo15} is a proof-of-work blockchain platform that initially claimed to implement a simplified version of a {\sc Ghost} consensus protocol, however, its current version differs significantly from the {\sc Ghost} protocol.
Ethereum does not take into account any stale block in the subtree when it decides the canonical path.
Instead, its branch selection process is solely based on the total difficulty value, consisting of the summation of difficulties of all the crypto puzzles of the block itself and all of its ancestors, as a weight of each branch.
Based on the difficulty value, Ethereum selects the branch with highest cumulated weight as a canonical branch.
Unlike the original {\sc Ghost} protocol, which treats each block equally in term of weight, a subtree with fewer number of blocks may be adopted as long as its total difficulty is higher than the others.

\subsection{Blockchain deployment environments} Blockchains are typically deployed in one of three commonly known environments depending on their access permissions: public, consortium, and private environments~\cite{buterin_public_2015}.
\begin{itemize}
\item {\bf Public blockchains} are the most permissive among all three; they are opened to any participant to access the systems.
As anyone is allowed to both read and write,
public blockchains generally rely on the Internet for communication.
\item {\bf Consortium blockchains} are more restrictive than public ones. Only a subset of the participating nodes may contribute to the consensus protocol that will lead to a new block being appended.
The read permissions in consortium blockchains could be either restricted to only the consensus participants or to all the public participants.
A good example for this kind of blockchain would be a consortium of financial institutions, who may compete with one another. 
To serve such purpose, consortium blockchains are usually deployed in environments that consist of multiple organization networks often interconnected by the Internet.
\item {\bf Fully private blockchains} are the most restrictive; the write permissions are preserved to only one organization. This type of blockchains could be deployed inthe private network of a single company, these block chains are often described as less interesting than the two other types of blockchains as they may not face the same security challenges.
\end{itemize}

\subsection{Mining pools and stratum servers} \emph{Mining pools} are groups of miners that combine their computational power to mine blocks and share the rewards among themselves. 
They are appealing in public blockchains as they allow miners to receive a smaller yet more frequent reward than if 
they were mining individually.
%
Each mining pool is viewed as a centralized miner from the rest of the system, however, their connectivity to the system is different from the connectivity of a central miner, as explained below.

At the heart of each pool is a small number of \emph{stratum servers}, which act as communication proxies between pool members and the rest of the blockchain network.
Information from the blockchain network flows in and out the mining pool via the stratum servers.
These servers coordinate the crypto-puzzle resolution by sending update and distributing workload to pool members.
This mechanism hides pool members behind the stratum servers, such that their information is not exposed to the blockchain network.
Finally, a stratum server hides information of a pool member from one another, as it eliminates the need for direct communication among the members.

\subsection{The Balance Attack}

The Balance Attack~\cite{NG17} is a recent theoretical generalization of the delay attack against proof-of-work blockchains; it relies on partitioning the network into subgroups of similar mining power  to achieve double spending with low mining power. 
It does not experiment how network attack can delay messages but rather focuses on the double spending risks when assuming network delays.
The attack is based on the fact that a proof-of-work blockchain favors partition tolerance rather than consistency, such that it still continues its usual operations with inconsistency and forkable chains of information.
The adversary splits the whole blockchain network into multiple subgroups with the aims of evenly balancing mining power of all subgroups, while the access to all of these subgroups are still protected from the adversary node.

The adversary can then perform double-spending by issuing conflicting transactions to many subgroups at once; without any communication among them, all the transactions will likely be accepted, because each subgroup is unaware that the same coins have been spent somewhere else.
The adversary could then influence the selection of the canonical chain among multiple branches by contributing his mining power to one of the subgroups. 
If the mining power is distributed evenly among all subgroups, the advantage of the adversary contribution makes one particular subgroup to possess higher mining power than the others and thus increases the chance for its branch to be adopted.
%

\begin{table*}
	\centering
	\caption{Existing top 10 Ethereum mining pools with mining power, stratum servers, location and AS numbers from July $27^{th}$ to Aug. $3^{rd}$}
	\small
	\resizebox{\textwidth}{!}{
		
		\label{table:miningpools}
		\begin{tabular}{|r|l|l|l|r|l|}
			\hline
			\multicolumn{1}{|c|}{\cellcolor[gray]{0.75} {Power (\%)}} & \multicolumn{1}{c|}{\cellcolor[gray]{0.75} {Pool Name}} & \multicolumn{1}{c|}{\cellcolor[gray]{0.75} {Stratum Servers}} & \multicolumn{1}{c|}{\cellcolor[gray]{0.75} {Location}} & \multicolumn{1}{c|}{\cellcolor[gray]{0.75} {ASN}} & \multicolumn{1}{c|}{\cellcolor[gray]{0.75} {AS Owner}}  \\ \hline \hline
			27.02                                     & f2pool                                  & eth.f2pool.com                                & Hangzhou, China                        & 37963                                   & Alibaba (China) Technology Co., Ltd.    \\ \hline
			\multirow{5}{*}{23.76}                    & \multirow{5}{*}{Ethermine}              & us1.ethermine.org                             & Montreal, Canada                       & 16276                                   & OVH SAS                                 \\ 
			&                                         & us2.ethermine.org                             & California, US                         & 63949                                   & Linode, LLC                             \\ 
			&                                         & eu1.ethermine.org                             & France                                 & 16276                                   & OVH SAS                                 \\ 
			&                                         & eu2.ethermine.org                             & France                                 & 16276                                   & OVH SAS                                 \\ 
			&                                         & asia1.ethermine.org                           & Singapore                              & 16276                                   & OVH SAS                                 \\ \hline
			\multirow{3}{*}{9.73}                     & \multirow{3}{*}{miningpoolhub}          & us-east.ethash-hub.miningpoolhub.com          & Georgia, US                            & 63949                                   & Linode, LLC                             \\ 
			&                                         & europe.ethash-hub.miningpoolhub.com           & Hesse, Germany                         & 63949                                   & Linode, LLC                             \\ 
			&                                         & asia.ethash-hub.miningpoolhub.com             & Tokyo, Japan                           & 63949                                   & Linode, LLC                             \\ \hline
			\multirow{5}{*}{9.7}                      & \multirow{5}{*}{Nanopool}               & eth-eu1.nanopool.org                          & France                                 & 16276                                   & OVH SAS                                 \\ 
			&                                         & eth-eu2.nanopool.org                          & France or Italy                        & 16276                                   & OVH SAS                                 \\ 
			&                                         & eth-asia1.nanopool.org                        & Singapore                              & 16276                                   & OVH SAS                                 \\ 
			&                                         & eth-us-east1.nanopool.org                     & Montreal, Canada                       & 16276                                   & OVH SAS                                 \\ 
			&                                         & eth-us-west1.nanopool.org                     & California, US                         & 20473                                   & Choopa, LLC                             \\ \hline
			\multirow{2}{*}{9.12}                     & \multirow{2}{*}{ethfans}                & guangdong-pool.ethfans.org                    & Fujian, China                          & 4134                                    & No.31,Jin-rong Street                   \\ 
			&                                         & huabei-pool.ethfans.org                       & Fujian, China                          & 4134                                    & No.31,Jin-rong Street                   \\ \hline
			\multirow{13}{*}{6.24}                    & \multirow{13}{*}{DwarfPool}             & eth-eu.dwarfpool.com                          & France                                 & 16276                                   & OVH SAS                                 \\ 
			&                                         & eth-us.dwarfpool.com                          & Montreal, Canada                       & 16276                                   & OVH SAS                                 \\ 
			&                                         & eth-us2.dwarfpool.com                         & Las Vegas, US                          & 53667                                   & FranTech Solutions                      \\ 
			&                                         & eth-ru.dwarfpool.com                          & France                                 & 16276                                   & OVH SAS                                 \\ 
			&                                         & eth-asia.dwarfpool.com                        & Taiwan                                 & 59253                                   & Leaseweb Asia Pacific pte. ltd.         \\ 
			&                                         & eth-cn.dwarfpool.com                          & Shanghai, China                        & 37963                                   & Alibaba (China) Technology Co., Ltd.    \\ 
			&                                         & eth-cn2.dwarfpool.com                         & Beijing, China                         & 37963                                   & Alibaba (China) Technology Co., Ltd.    \\ 
			&                                         & eth-sg.dwarfpool.com                          & Singapore                              & 59253                                   & Leaseweb Asia Pacific pte. ltd.         \\ 
			&                                         & eth-au.dwarfpool.com                          & Melbourne, Australia                   & 38880                                   & Micron21 Melbourne Australia Datacentre \\ 
			&                                         & eth-ru2.dwarfpool.com                         & Moscow, Russia                         & 42632                                   & MnogoByte LLC                           \\ 
			&                                         & eth-hk.dwarfpool.com                          & Hong Kong                              & 45102                                   & Alibaba (China) Technology Co., Ltd.    \\ 
			&                                         & eth-br.dwarfpool.com                          & Sao Paulo, Brazil                      & 262287                                  & Maxihost Hospedagem de Sites Ltda       \\ 
			&                                         & eth-ar.dwarfpool.com                          & Rosario, Argentina                     & 27823                                   & Dattatec.com                            \\ \hline
			4.45                                      & BW                                      & ether.bw.com                                  & Wuhan, China                           & 58563                                   & CHINANET Hubei province network         \\ \hline
			\multirow{4}{*}{3.34}                     & \multirow{4}{*}{Ethpool}                & us1.ethpool.org                               & Montreal, Canada                       & 16276                                   & OVH SAS                                 \\ 
			&                                         & us2.ethpool.org                               & Montreal, Canada                       & 16276                                   & OVH SAS                                 \\ 
			&                                         & eu1.ethpool.org                               & France                                 & 16276                                   & OVH SAS                                 \\ 
			&                                         & asia1.ethpool.org                             & Singapore                              & 16276                                   & OVH SAS                                 \\ \hline
			1.83                                      & Coinotron                               & coinotron.com                                 & Poland                                 & 51290                                   & HOSTEAM-AS                              \\ \hline
			0.88                                      & Poolgpu                                 & eth.poolgpu.com                               & Hangzhou, China                        & 37963                                   & Alibaba (China) Technology Co., Ltd.    \\ \hline\hline
			\cellcolor[gray]{0.75} 96.07  &  \multicolumn{5}{l|}{\cellcolor[gray]{0.75} Total} \\ \hline
		\end{tabular}
	}
\end{table*}

\section{Man-in-the-middle Double Spending in the public Ethereum blockchain}
\label{sec:problem}

In this section, we show experimentally how someone can supposedly double spend by partitioning Ethereum and then 
explain why the network connectivity makes the success of this attack almost impossible.

In order to gain some insights regarding 
the public Ethereum blockchain
we combined the name and block contributions of the top-10 mining pools as observed on \texttt{\url{http://etherscan.io}} during one week on August $3^{rd}$, 2017
with their network connectivity information in \tablename~\ref{table:miningpools}. To this end, to each named mining pool we registered a miner that could 
gather IP and \emph{Autonomous System (AS)} information.
ASes are groups of networks under control of a single technical administration~\cite{hawkinson_guidelines_1996}.
%
%
In particular, we estimated locations of the servers by querying 5 geo-IP databases~\cite{noauthor_db-ip_nodate,noauthor_ip_nodate-1,noauthor_ip_nodate-2,noauthor_ip_nodate,noauthor_ip_nodate-3}.
To reduce the inaccuracies of geo-locations, we extracted the location indicated in the majority of these databases.
%
To retrieve the number and owner of each AS 
we relied on~\cite{noauthor_caida:_nodate} and \cite{noauthor_merit_nodate}, both sources are based on the \texttt{whois} service.
ASes have their own routing policy for internal traffic but use \emph{Border Gateway Protocol (BGP)}~\cite{hares_border_2006} for dynamic inter-AS routing.

Unfortunately, BGP does not incorporate a mechanism to check whether an origin AS owns the IP prefixes that it announces. This makes a protocol vulnerable to route hijacking.


\subsection{Double spending is easy in case of route hijacking}\label{sec:bgp}
\begin{figure}
	\centering
	\includegraphics[width=0.45\textwidth]{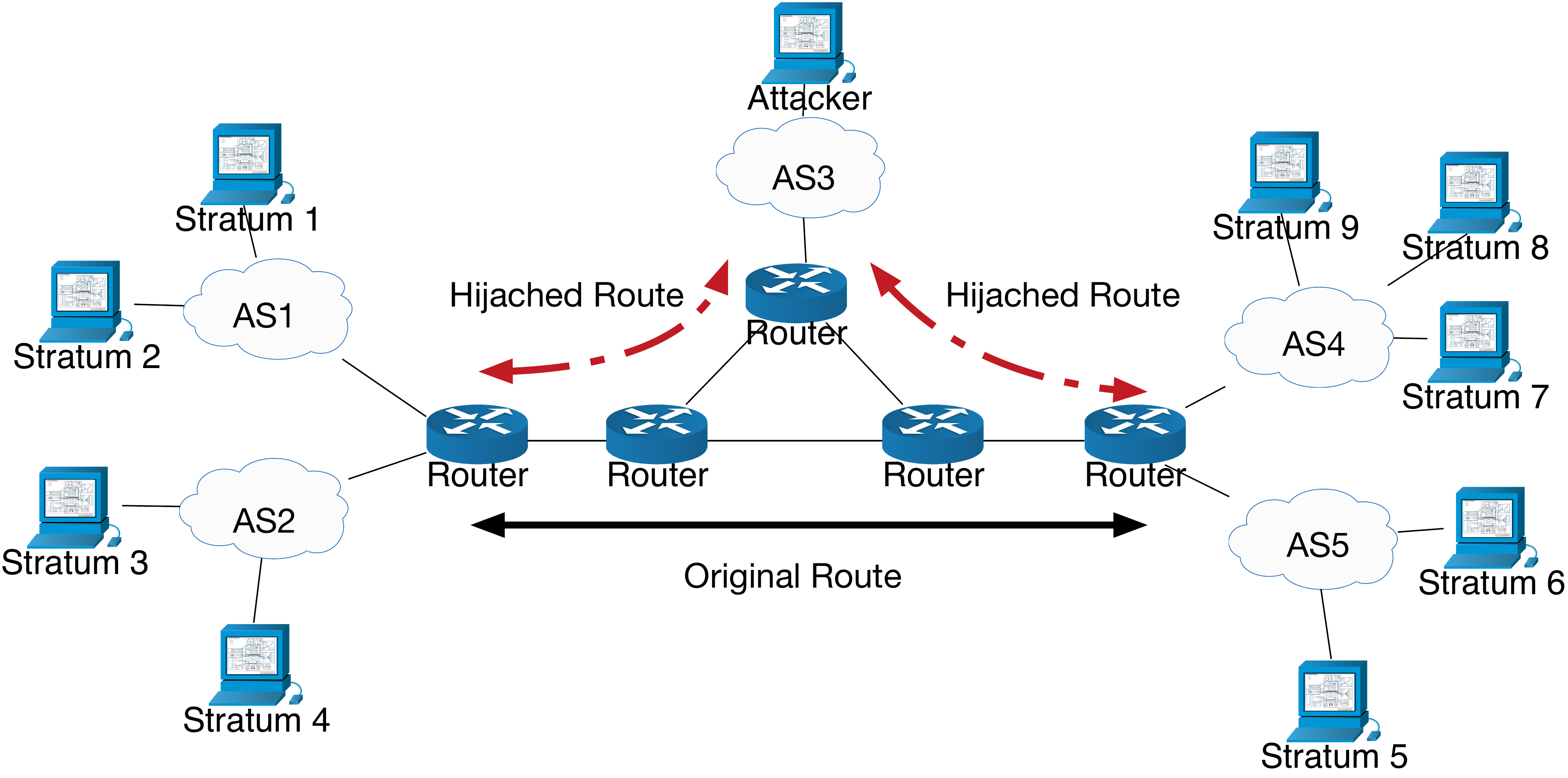}
	\caption{Experimental topology for public network}
	\label{fig:TopoBGP}
	\vspace{-.4cm}
\end{figure}

To quantify the risk of a partitioning attack, we emulated the aforementioned Ethereum top-10 mining pools by deploying 10 virtual machines (VMs) linked through 5 BGP routers, as shown in \figurename~\ref{fig:TopoBGP}, and 
controlled in our private cloud infrastructure via OpenStack.
%
To obtain the mining power distribution of mining pools retrieved in \tablename~\ref{table:miningpools} among our own VMs, we fixed the quantum of CPU time allocated to each machine using Linux \texttt{cgroups}~\cite{heo_control_2015}.
\texttt{cgroups} allow us to specify the CPU quota $Q$ that a VM can consume within a period of time $T$.
Given the same value of $T$, we vary $Q$ on all virtual machines based on their correspondent mining power percentage of the pools.
As a result, we obtained the proportion we listed in \tablename~\ref{table:miningpools} close to 1 decimal.

We then combined a BGP-hijacking attack with the balance attack~\cite{NG17} to evaluate the risks of double spending in Ethereum v1.5.
First, the BGP hijacking is used to delay communication, then the balance attack is used to turn these delays into double spending.
To this end, we assign the role of the adversary to one of the mining pool in each of our attack instances.
As indicated in \figurename~\ref{fig:TopoBGP}, the adversary takes control over one BGP router to prevent AS1 and AS2 from communicating with AS4 and AS5 during 7 minutes.
During that time, the adversary issues a transaction to one group and contributes to the block creation of the other group in order to discard its previously issued transaction.
Since it is commonly recommended to wait for 12 confirmations to be confident about the immutability of a transaction since the version Homestead of Ethereum, we consider 
the double spending successful when the transactions contained in a block followed by 11 consecutive blocks gets discarded. 
%

After the 7 minutes communication delay, we observed whether the adversary transaction is discarded due to the choice of the canonical chain in 30 consecutive runs and concludes upon the average success of the attack.
%
As indicated in \tablename~\ref{table:miningpools2}, we observe that only 10\% of the mining power is sufficient for the double spending to be successful most of the time.
With only 27\% of the mining power, the success of the attack reaches 76\%.

This result confirms the claim from the literature that it is supposedly feasible to attack public blockchains~\cite{AZV17}, however, this is without taking into account the nature of the network topology. 
We explain in Section~\ref{sec:difficult-attack} why the topology described in \tablename~\ref{table:miningpools} makes the attack of the public Ethereum blockchain almost impossible.

\subsection{Partitioning Ethereum mining pools turns out to be hard}\label{sec:difficult-attack}

While we showed that double spending could be easily achieved by partitioning public blockchains, it turns out that partitioning the mining power of Ethereum mining pools is almost impossible.
\begin{figure}
	\centering
	\includegraphics[width=0.45\textwidth]{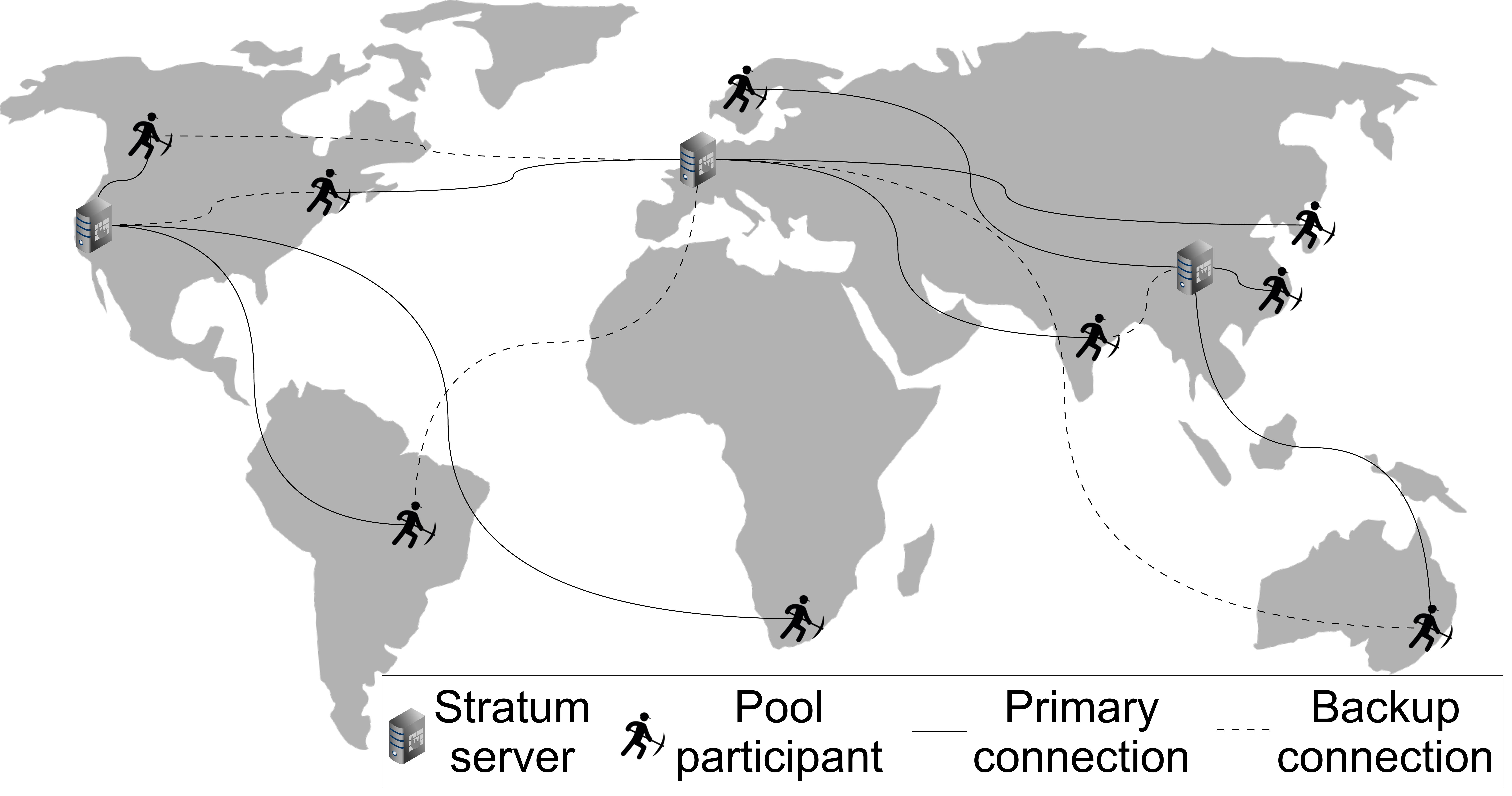}
	\caption{Hypothetical stratum servers and pool participants}
	\label{fig:stratum_and_mining_pools}
\end{figure}

\tablename~\ref{table:miningpools} lists the stratum servers of the top-10 Ethereum mining pools we retrieved.
We noticed experimentally that if one of the stratum server become unresponsive, then the corresponding miners would connect to the next stratum server they operate in order to remain connected to the pool.
Hence, partitioning may result in having miners reconnect to a different AS.
As an example, consider  \figurename~\ref{fig:stratum_and_mining_pools}, where a miner in India primarily connected to Europe may reconnect to China.

In addition, it is more difficult to determine the precise proportion of mining power connected to each stratum server, again due to the numerous stratum servers each miner operates.
%
Indeed, a mining pool identifier is nothing more than the wallet address to receive reward when a pool successfully mines a block.
While it is possible to determine a block miner by examining header information, there is no way to pin down to the stratum server, as long as these servers put their reward into the same wallet address.

\begin{table*}
	\centering
	\caption{Success of double spending with mining pools of similar power 
	to the top 10 Ethereum mining pools from July $27^{th}$ to Aug. $3^{rd}$}
	\small
	\resizebox{\textwidth}{!}{
		\label{table:miningpools2}
		\setlength{\tabcolsep}{15pt}
		\begin{tabular}{|c|l|l|c|c|}
			\hline
			\rowcolor[HTML]{C0C0C0} 
			\multicolumn{2}{|c|}{\cellcolor[HTML]{C0C0C0}\begin{tabular}[c]{@{}c@{}}Mining power of members \\ in an adversary subgroup (\%)\end{tabular}} & \multicolumn{1}{c|}{\cellcolor[HTML]{C0C0C0}}                                                                                                               & \cellcolor[HTML]{C0C0C0}                                                                                              & \cellcolor[HTML]{C0C0C0}                                           \\ \cline{1-2}
			\rowcolor[HTML]{C0C0C0} 
			Adversary                                  & \multicolumn{1}{c|}{\cellcolor[HTML]{C0C0C0}The rest of subgroup}                                 & \multicolumn{1}{c|}{\multirow{-2}{*}{\cellcolor[HTML]{C0C0C0}\begin{tabular}[c]{@{}c@{}}Mining power of members \\ in a victim subgroup (\%)\end{tabular}}} & \multirow{-2}{*}{\cellcolor[HTML]{C0C0C0}\begin{tabular}[c]{@{}c@{}}Difference between \\ two subgroups\end{tabular}} & \multirow{-2}{*}{\cellcolor[HTML]{C0C0C0}\begin{tabular}[c]{@{}c@{}}Double Spending \\Success rate (\%)\end{tabular}} \\ \hline
			27.02     & 23.76, 6.24, 3.34, 0.88                                           & 9.73, 9.7, 9.12, 4.45, 1.83                                                    & 26.41                                                                                                                 & 76.7                                                               \\ 
			23.76     & 27.02, 6.24, 1.83, 0.88                                           & 9.73, 9.7, 9.12, 4.45, 3.34                                                    & 23.39                                                                                                                 & 63.3                                                               \\ 
			9.73      & 27.02, 9.12, 4.45, 1.83, 0.88                                     & 23.76, 9.7, 6.24, 3.34                                                         & 9.99                                                                                                                  & 56.7                                                               \\ 
			9.70       & 27.02, 9.12, 4.45, 1.83, 0.88                                     & 23.76, 9.73, 6.24, 3.34                                                        & 9.93                                                                                                                  & 43.3                                                               \\ 
			9.12      & 27.02, 9.7, 4.45, 1.83, 0.88                                      & 23.76, 9.73, 6.24, 3.34                                                        & 9.93                                                                                                                  & 43.3                                                               \\ 
			6.24      & 27.02, 9.7, 4.45, 3.34                                            & 23.76, 9.73, 9.12, 1.83, 0.88                                                  & 5.43                                                                                                                  & 40                                                                 \\ 
			4.45      & 27.02, 9.7, 6.24, 1.83, 0.88                                      & 23.76, 9.73, 9.12, 3.34                                                        & 4.17                                                                                                                  & 40                                                                 \\ 
			3.34      & 27.02, 9.7, 6.24, 1.83, 0.88                                      & 23.76, 9.73, 9.12, 4.45                                                        & 1.95                                                                                                                  & 43.3                                                               \\ 
			1.83      & 27.02, 9.7, 6.24, 3.34, 0.88                                      & 23.76, 9.73, 9.12, 4.45                                                        & 1.95                                                                                                                  & 33.3                                                               \\ 
			0.88      & 27.02, 9.7, 6.24, 3.34, 1.83                                      & 23.76, 9.73, 9.12, 4.45                                                        & 1.95                                                                                                                  & 26.7                                                               \\ \hline
		\end{tabular}
	}
\end{table*}
Second, the stratum servers typically hide the location of the mining pool participants, which makes it hard to isolate a group of pools of a specific mining power.
In particular and as described in \figurename~\ref{fig:stratum_and_mining_pools}, one cannot prevent a miner from India to reconnect to a stratum node located in China.
Without information about the miners for a stratum server, 
one cannot guarantee the partition success of a network attack.
It may \textit{(i)}~isolate a stratum server along with its miners completely, \textit{(ii)}~partition some miners, which reduces only a fraction of computational power from the pool, or \textit{(iii)}~cut off the connectivity between a stratum server and pool participants, such that those participants decide to reconnect to different stratum servers.

\begin{figure}
	\centering
	\includegraphics[width=0.46\textwidth]{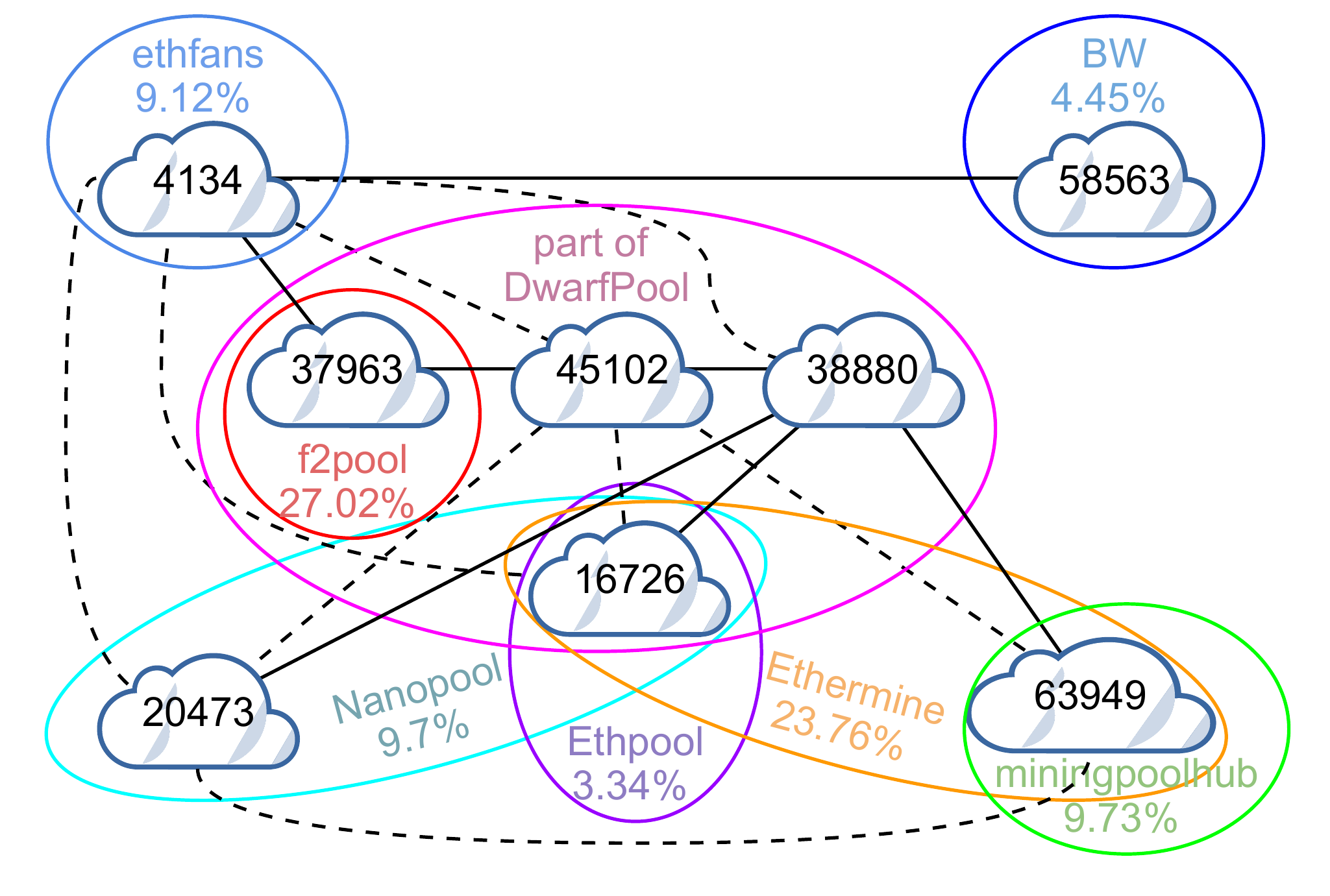}
	\caption{The inter-AS connections among the group of ASes that host stratum servers for public Ethereum mining pools. Each cloud represents an AS. An oval shape represents a mining pool. A thick line illustrates a link between two adjacent ASes or a peering connection, while a dash line indicates the existence of an indirect path between two ASes that requires only one transit AS in the middle.}
	\label{fig:asandpools}
\end{figure}

Third, BGP-hijacking cannot affect the direct interconnection between ASes, because 
ASes are aware of static network prefixes that belong to their peer ASes.
Apart from exchanging routes at the \emph{Internet Exchange Points (IXPs)}, any pair of ASes may decide to establish either layer 2 or layer 3 links to connect their networks directly.
This
prevents dynamic routing attacks like the BGP hijacking we discussed above in Section~\ref{sec:bgp}.
To better understand the applicability of the attack to the Ethereum public blockchain, we retrieved the direct peering information of the 8 ASes we identified using available information~\cite{as_caida:_2017} and listed these interconnections in \figurename~\ref{fig:asandpools}.
Among the top-10 public Ethereum mining pools, 7 of them solely rely on this group of ASes; together, they account for more than 87\% from the total mining power of the network.
As the majority of ASes in this group are linked by direct peering, it appears extremely difficult to partition Ethereum's overlay.
For example, \texttt{f2pool} may send and update to \texttt{ethfans} via a peering connection, which in turn forwards the update to \texttt{BW} via another peering connection.
Without an adversary gaining access to configuration on the border routers of these ASes, it will remain difficult to partition a pool from the rest of the group.

\section{Man-in-the-middle Double Spending in a Consortium Ethereum Blockchain}
\label{sec:consortium}

As opposed to the previous section, we now focus on a consortium context and show how easy it is to double spend in Ethereum.
Since participants can be located in different regions around the globe, the members of a consortium typically use the Internet and multiple ASes for communicating.
%
For the sake of simplicity, let us consider that all members have an equal amount of mining power.
\footnote{A skewed distribution of mining power among members makes the attack easier as quantified theoretically~\cite{NG17}.}
Although the consortium may include competitors, these are usually not incentivized to participate based on a mining rewards. 
This is why there is generally no mining pools in consortia and it is quite easy to double spend as we illustrate below.

\subsection{Setting up a double spending attack with BGP-hijacking}

In order to quantify the risks of one member of the consortium to steal assets from its competitors, we deployed a testnet
of 9 VMs: one adversary and two groups of similar mining power of 4 VMs each.
All VMs run go-ethereum or \texttt{geth} version 1.5 (Ethereum Homestead) and are connected through a network similar to the previous experiments
configured with different message latency to mimic multiple realistic geographical scales: from almost no delay to 250\,ms delay.
We setup an Ethereum overlay so that every node is connected to one another through a fully mesh logical topology that we manually configured.


To provide connectivity among the nodes, 5 routers are used similar to the configuration shown in 
\figurename~\ref{fig:TopoBGP} where stratum server 8 is removed and where the remaining stratum servers are replaced by miners.
Each router employs a Quagga BGP daemon~\cite{noauthor_quagga_nodate} that advertises reachability to the adjacent networks to its neighbors.
By repeating this process, 
each router is eventually informed about the location of any destination network of some AS.
Upon reception of the advertisement, the BGP process running in each router uses the information of all these advertised paths to update its routing table.

\begin{figure}
	\centering
	\subfigure[Difficulty Evolution \label{fig:difficultyout}]{\includegraphics[width=0.24\textwidth]{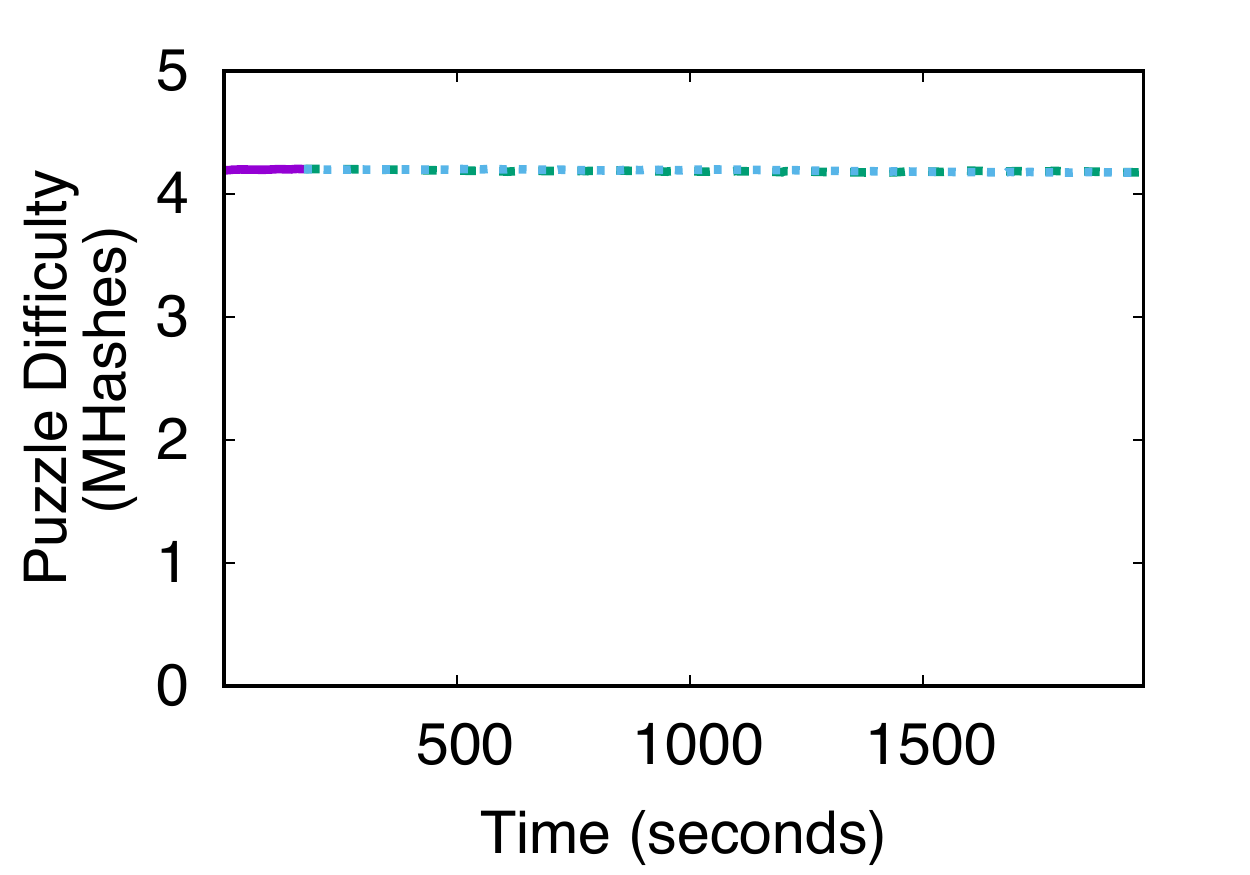}}
	\subfigure[Difficulty Evolution Zoomed around 4.2 MH\label{fig:difficultyin}]{\includegraphics[width=0.24\textwidth]{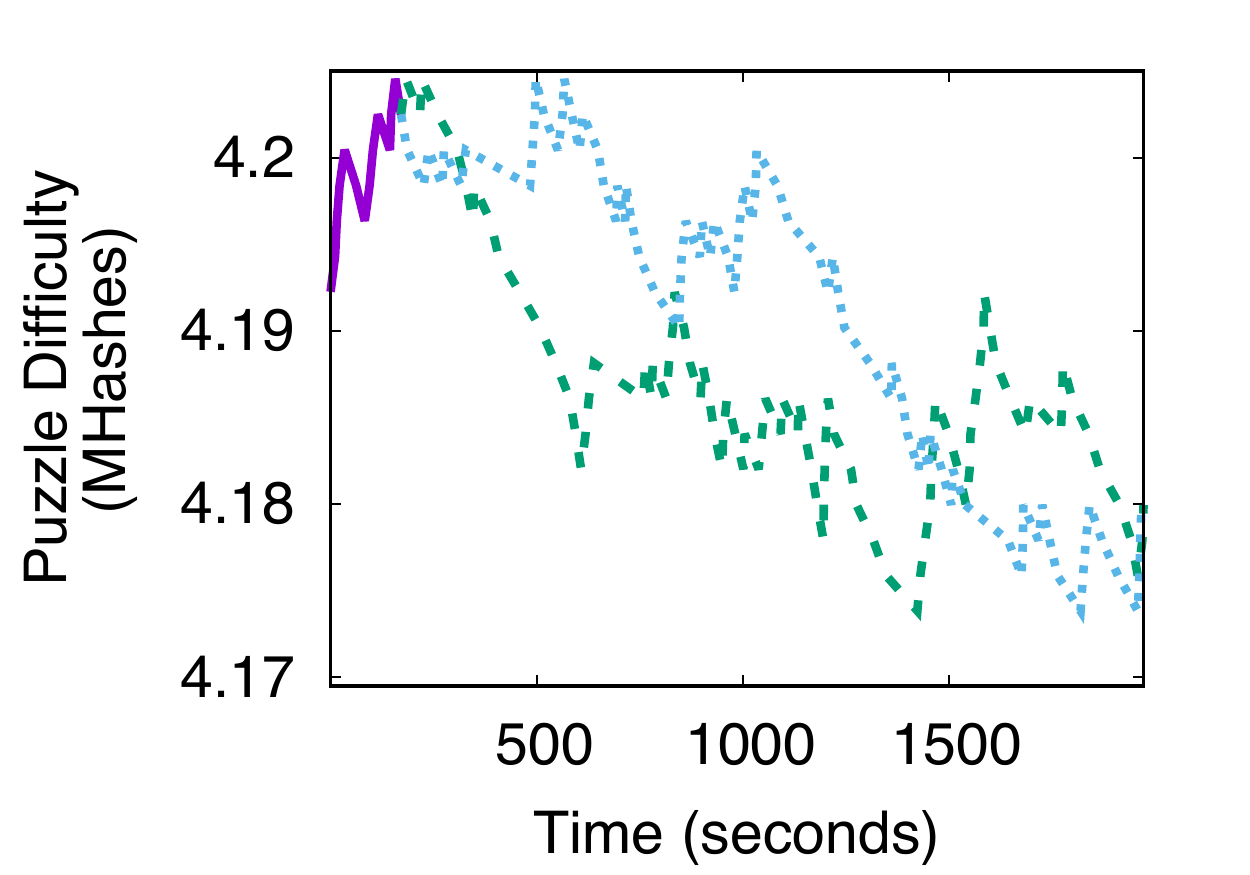}}
	\subfigure{\includegraphics[width=0.15\textwidth]{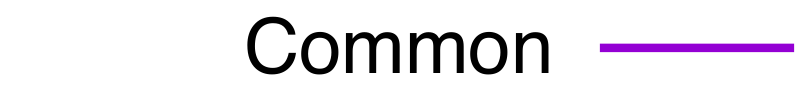}}
	\subfigure{\includegraphics[width=0.15\textwidth]{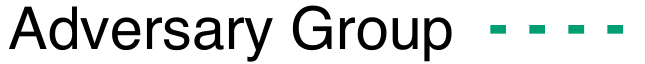}}
	\subfigure{\includegraphics[width=0.15\textwidth]{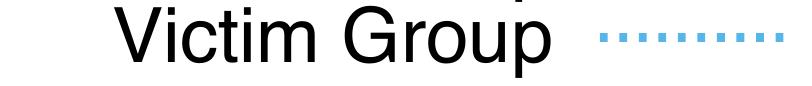}}
	\caption{Difficulty chart during a 30 minute network partitioning}
	\label{fig:difficulty}
\end{figure}

We now described how one node selected as the adversary can exploits BGP hijacking to get full control of a router and double spend.
Initially, all BGP routers are properly configured, where 
communication of nodes between two legitimate ASes are routed via either 4 or 2 BGP routers, depending on the location of the AS. 
To start the attack, the adversary first configures a router under its control to maliciously advertise, to the neighbor routers, non-existing direct routes to those two ASes.

Since there is no mechanism to check the validity of all networks in an advertisement message, the neighbor routers will simply accept such routes as a route modification and update their routing tables.
This adds the malicious router as an extra hop in the middle of two ASs, which create the path of 5 BGP routers instead of 4 for both westbound and eastbound communications.
Next, the adversary blocks or drops a certain type of the traffic (Ethereum traffic in this case), while the adversary still holds the capability to talk with both ASs.
As a consequence, the Ethereum blockchain is partitioned into two subgroups with the adversary in the middle of the communication path.

To exploit this partition to double spend, 
the adversary then proceed as in the Balance Attack~\cite{NG17} by issuing two conflicting transactions, each of them to one subgroup.
Each of the transactions transfers more than half of the coins available in the wallet, so that they are clearly in conflict.
After that, the adversary waits long enough for a transaction on the victim side to get committed, which results in waiting for a block to include the transaction and the observation that a sufficient number of subsequence blocks were mined.
Finally, after a sufficiently long period of time, the adversary simply stops BGP hijacking and lets the two Ethereum subgroups reconnect.
When the network is no longer partitioned, the Ethereum protocol selects the branch that does not contain the transaction included in the blockchain by the victim subgroup.
Any real goods purchased with this transaction will be owned by the adversary despite leaving no trace of this transaction or purchase.

\subsection{Success of double spending after a BGP-hijacking attack}

As mentioned in Section~\ref{sec:background}, Ethereum adjusts the difficulty of their crypto puzzles on every block; our experiments reveal that, however, the adaptation does not reach convergence fast enough to reflect the change in mining power of the network.
\figurename~\ref{fig:difficulty} shows the variation of the crypto puzzle difficulty over time, both before and during the man-in-the-middle attack.
During the network partition, the mining power on an adversary subgroup and victim subgroup are 55.6\% and 44.4\% respectively.
Looking closely at the difficulty trend in \figurename~\ref{fig:difficultyin}, we can observe that the values fluctuate during some time, while the overall trend is going downward.
One may imagine the difficulty on the adversary side to be much higher than the victim side, because of the about 11\% higher mining power, but this is also not the case here.
Furthermore, one could expect the difficult to drop to roughly 2.1 Million Hashes (MH), which is about half of its previous value; the difficulty value, however, still remains higher than 4.17 MH even after 30 minutes.


\begin{table}
	\centering
	\caption{Double Spending Success rate with BGP hijacking in a Consortium Deployment}
	\label{tab:bgpsuccessrate}
	\resizebox{.47\textwidth}{!}{
		\begin{tabular}{lcccccccccc}
			\toprule
			& \multicolumn{10}{c}{Attack Duration (min)} \\
			Network Delay  &  3 & 4 & 5 &6 & 7 & 8 & 9 & 10 & 11 & 12 \\
			\midrule
			<1 ms &  3 \% & 23 \% & 57 \% & 60 \% & 53 \% & 60 \% & 77 \% & 63 \% & 80 \% & 77 \%\\
			250 ms & 10 \% & 30 \% & 47 \% & 47 \% & 63 \% & 50 \% & 53 \% & 43 \% & 60 \% & 63 \%\\
			\bottomrule
	\end{tabular}}
\end{table}

\begin{figure}
	\centering
	\includegraphics[width=.35\textwidth]{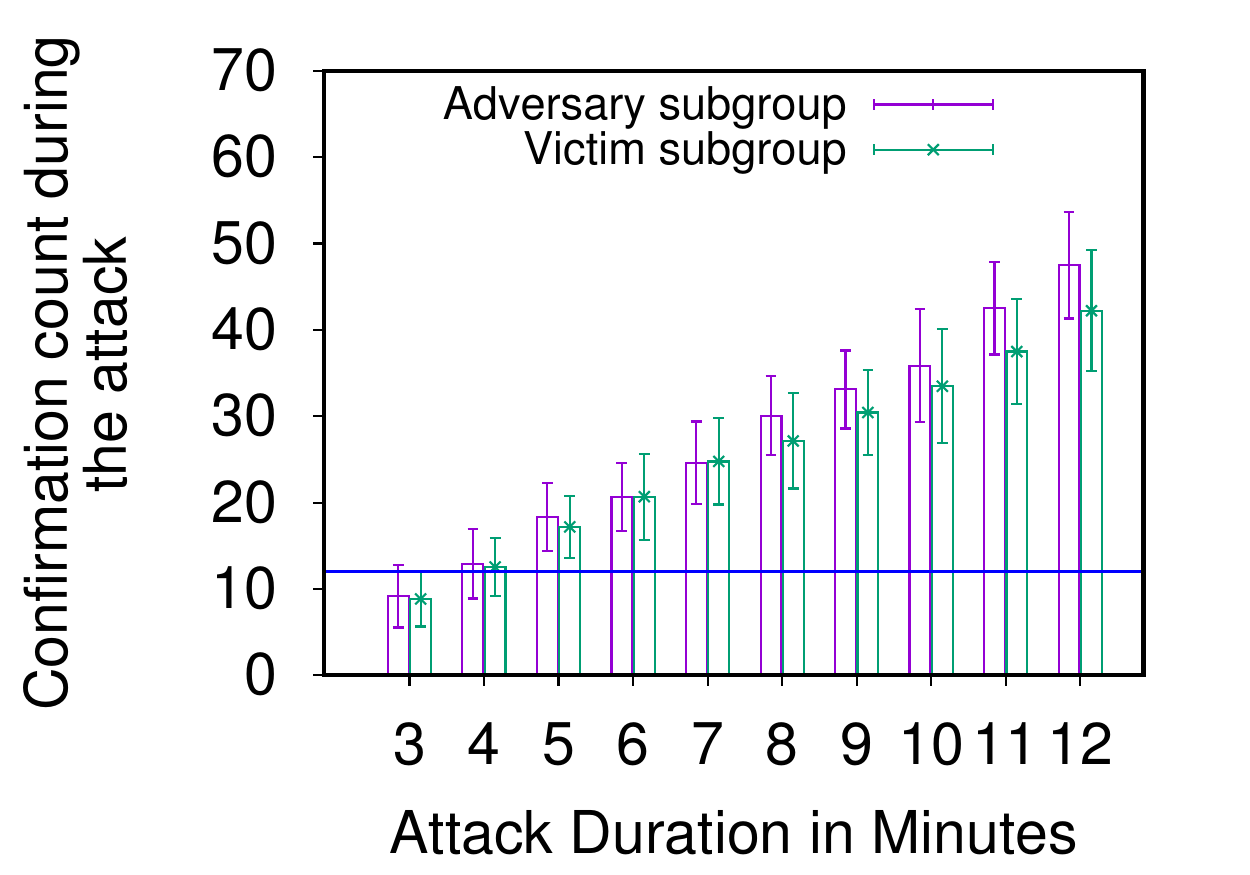}
	\caption{Block Count during the BGP hijacking based on duration of network segmentation}
	\label{fig:blocks-in-bgp}
\end{figure}

\tablename~\ref{tab:bgpsuccessrate} shows the average number of successful attempts of double spending from 30 trials for each attack duration, while \figurename~\ref{fig:blocks-in-bgp} shows the average number of block mined by the adversary subgroup versus the victim subgroup. 
The blue horizontal line in \figurename~\ref{fig:blocks-in-bgp} represents the baseline of 12 confirmations, in other word, the transaction has obtained enough subsequent blocks to be considered committed.
\tablename~\ref{tab:bgpsuccessrate} indicates that a 3-minute attack duration hardly yields a successful double spending, this is because the time is too short for the miners on each subgroup to mine enough blocks to commit (or obtain sufficient confirmations for) a transaction.
Results from this experiment illustrate that the adversary may perform double spending with high chance of success, such as more than 75\% with an attack lasting 9 minutes.
We can also see in the \figurename~\ref{fig:blocks-in-bgp} that, as expected, the longer the attack duration, the high number of block created during the attack. 


\begin{figure}
	\centering
	\includegraphics[width=.35\textwidth]{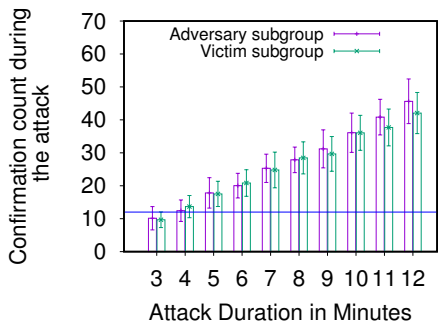}
	\caption{Block Count during the BGP hijacking based on duration of network segmentation with 250ms delay}
	\label{fig:blocks-in-bgp-delay}
\end{figure}

We then perform a similar experiment while artificially introducing  250\,ms of delay between two ASes on the left and on the right of the network topology.
Similarly to the previous experiment, \tablename~\ref{tab:bgpsuccessrate} and \figurename~\ref{fig:blocks-in-bgp-delay} present the average number of successful attempts from 30 trials for each attack duration and the average number of blocks mined by two subgroups respectively.
We can see that both adversary and victim subgroups produce less blocks than in the first set of experiments without delay.
It results in a lower probability of double spending success on average.
This may be due to fact that delays slow down the block propagation, thus the system is likely to mine more stale blocks, which in turn reduces the success rate.


To summarize, our experiments exhibited several trends.
First, we saw that the attack duration contributes to the success of double spending, both with or without artificial delays.
After a certain attack duration, the success no longer increases significantly, as can be seen by the slight success increase between 9 and 12 minutes.
Second, from the comparison between the experiment with and without delays, we can see that the delays among peers in the topology directly affects the probability of the attack.
The higher the delay of the network, the lower the chance of successful attack.
Finally, it is important to note that even though double spending did not succeed in all our experiments, we were able to successfully perform BGP hijacking successfully.

\section{Man-in-the-middle Double Spending in a Private Ethereum Blockchain}
\label{sec:private}

In this section, we 
show the feasibility of double spending on the private Ethereum blockchain using another man-in-the-middle attack.
For the purpose of this work, we define a private blockchain as a group of Ethereum miners peering within the same AS via a local area network (LAN) with low latency. 
Note that this is reasonable to use an Ethernet network with a single layer 2 broadcast domain in the private blockchain context.
Considering that a minimum network security can be in place in these network, the adversary may exploit an ARP-spoofing attack to take the control of the network without the need for any configuration change on network elements.
As there is less demand to compete among the miners in this setting, we will assume that all of the nodes possess the same amount of computing resources.

\subsection{Setting up of a double spending attack with ARP spoofing}

\begin{figure}
	\centering
	\includegraphics[width=0.3\textwidth]{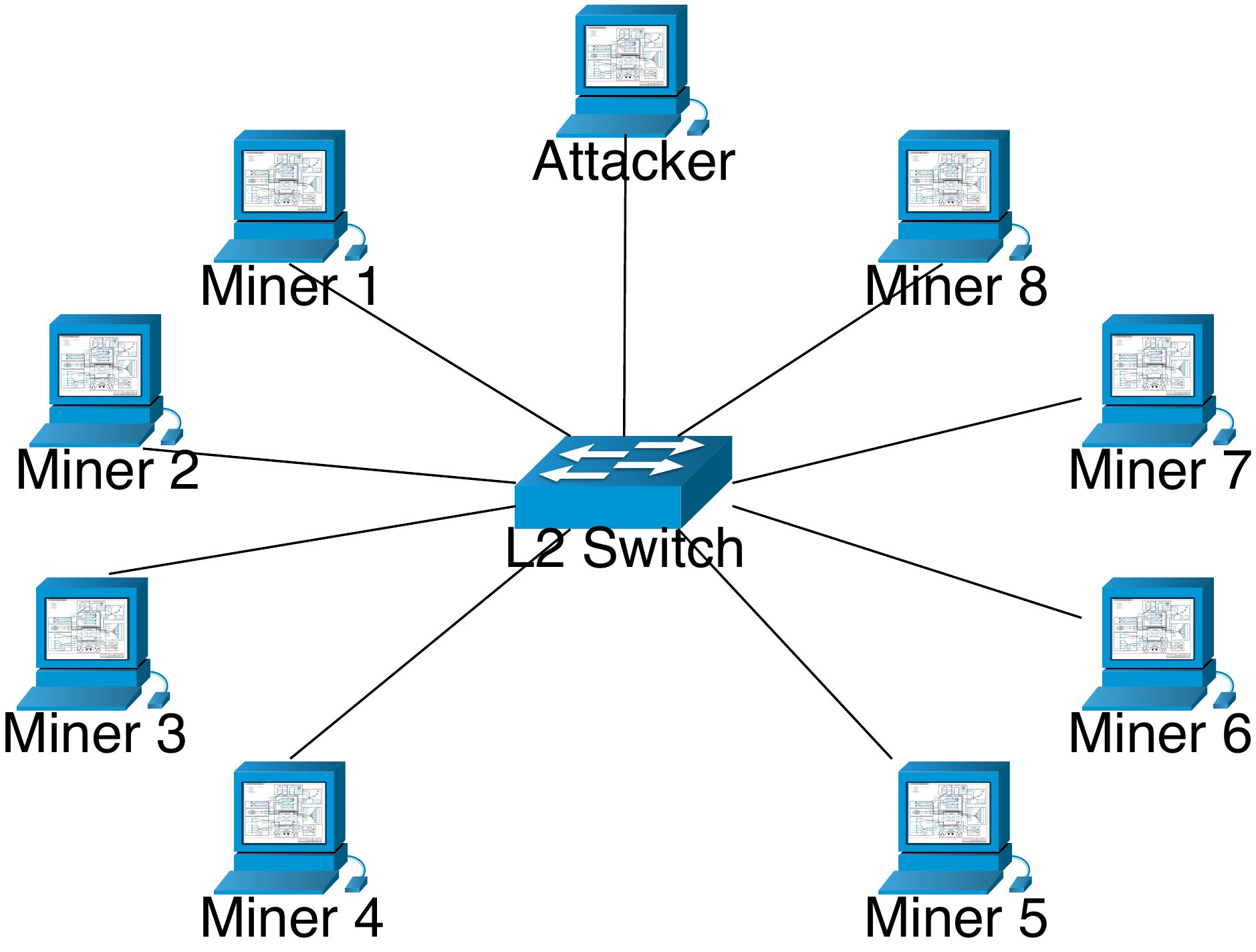}
	\caption{Experimental Topology for Local Area Network}
	\label{fig:TopoLAN}
\end{figure}

To quantify the chance of successful double spending attack on Ethereum in a private blockchain context, we run the experiment on our private cloud infrastructure with a low latency network.
The setup is similar to the previous experiment for a consortium context, except that all of the virtual machines are linked together via a physical Ethernet switch.
As a result, we have deployed a topology as shown in \figurename~\ref{fig:TopoLAN}.

In the private blockchain context, our adversary employs ARP spoofing technique~\cite{ramachandran_detecting_2005} to perform a man-in-the-middle attack.
Since there is no authentication mechanism used in a common ARP protocol, the adversary can send IP packets with fabricated MAC address.
Considering a simple communication between node A and B, the adversary can tell node A that node B's IP address is mapped to its MAC address and vice versa.
Any packet from node sent to node B then route through the adversary first; as a consequence, the adversary may disrupt the communication by delaying the network traffic.
By positioning itself in the middle of a communication path among all of the minders, the adversary separates the blockchain network into two subgroups, while they still hold the capability to communicate with both of them.
Under this circumstance, the adversary may issue two conflicting transactions to perform double spending similar to what we explained in Section~\ref{sec:consortium}.

\subsection{Success of double spending after an ARP spoofing attack}

\begin{table}
	\centering
	\caption{Double spending success rate with ARP spoofing}
	\label{tab:arpsuccessrate}
	\resizebox{.47\textwidth}{!}{
		\begin{tabular}{lcccccccccc}
			\toprule
			& \multicolumn{10}{c}{Attack Duration (min)} \\
			Network Delay  &  3 & 4 & 5 &6 & 7 & 8 & 9 & 10 & 11 & 12 \\
			\midrule
			<1 ms &  7 \% & 23 \% & 47 \% & 53 \% & 57 \% & 67 \% & 47 \% & 53 \% & 67 \% & 80 \%\\
			\bottomrule
	\end{tabular}}
\end{table}

\begin{figure}
	\centering
	\includegraphics[width=.35\textwidth]{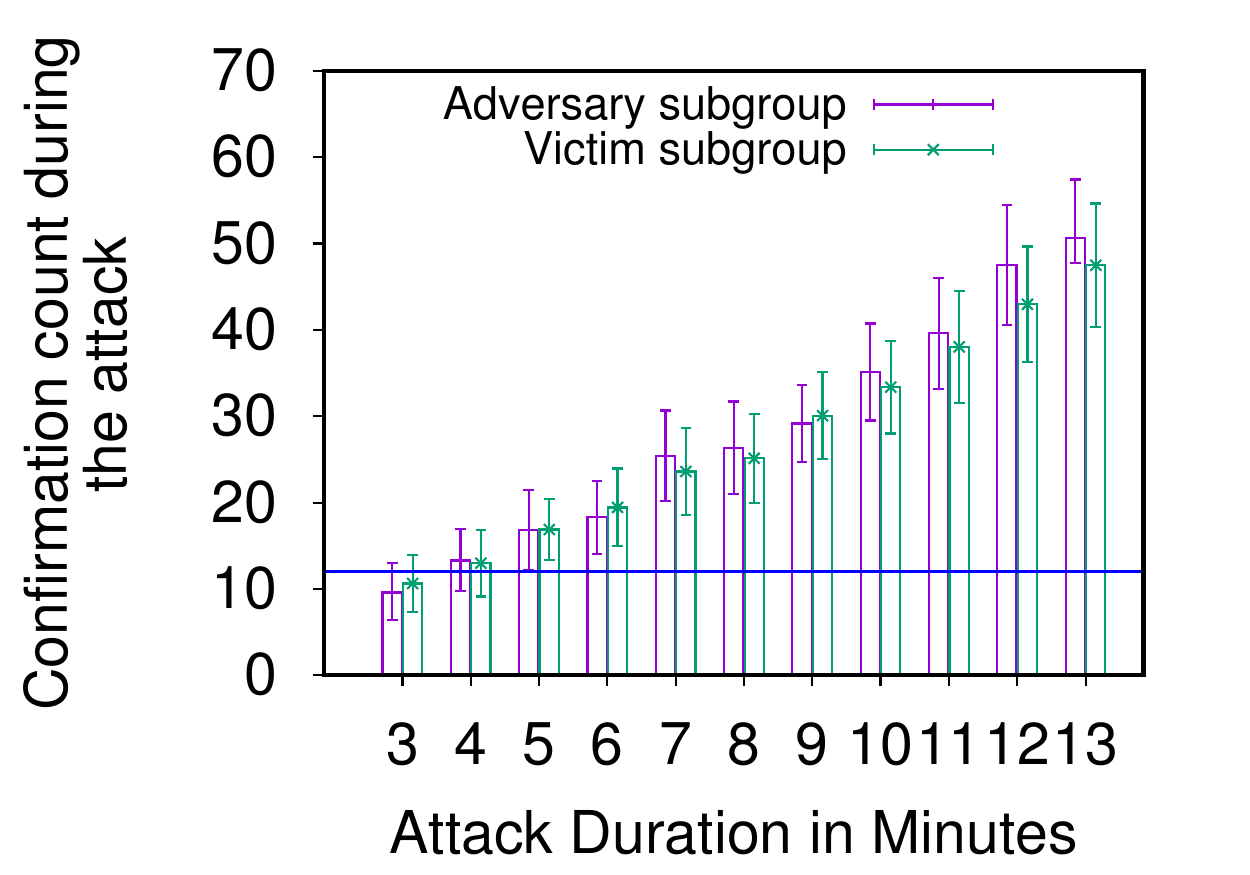}
	\caption{Block Count during the ARP Attack based on duration of network segmentation}
	\label{fig:blocks-in-arp}
\end{figure}

Results obtained from experiments in a private context exhibit similar behaviors to the consortium environment.
\tablename~\ref{tab:arpsuccessrate} shows the average number of successful double spending from 30 runs in function of the attack duration, while \figurename~\ref{fig:blocks-in-arp} shows the average number of block mined by both adversary and victim subgroups. 
In this context, we observed as expected that the longer the attack duration, the higher number of block mined by both adversary and victim subgroups.
We observed that the double spending success rate reaches its highest probability at 80\% with 12 minutes attack duration.
The overall trend is similar to a consortium blockchain, however, we observe a drop for an attack during 9 minutes.
While we cannot clearly explain this event, we hypothesize that is could be due to the randomness of block mining in Ethereum.

In our experiment, the ARP-spoofing attack always succeeded (i.e., the network was successfully partitioned), however there are shortcomings that decrease its of chance of success in practice.
By contrast with BGP hijacking, the adversary doing an ARP spoofing does not require any control of network entity.
The adversary node needs, however, to bear the burden of all communication traffic among the miners in the blockchain network, hence causing an additional overhead of the adversary.
Further, the risk of ARP-spoofing attack can be greatly reduced with a network protection feature within the switches.
In particular, most modern network switches have capability to detect and restrict the number of MAC addresses per a network port, which in turn prevents ARP-spoofing; 
such a feature is usually turned on by default on non-consumer hardware as well as in virtualized environments.

Overall, double spending with an ARP spoofing attack shows a similar trend as with a BGP hijacking attack.
While the local network was successfully partitioned, a low attack duration, such as 3 minutes, hardly results in a successful double spending.
The probability of the attack increases to certain degree as the attack duration lasts longer.
Similarly, the increase in success rate eventually reaches a plateau after some attack duration is reached.

\section{Discussion}
\label{sec:discussion}

In this section, we discuss the consequences of double spending using man-in-the-middle attacks against Ethereum blockchain, as well as potential countermeasures.
To begin with, we want to quantify the attack impact depending on the time the adversary runs the attack continuously, as this  could happen in a real-world blockchain situation where a system is not under constant monitoring.
Later, we present three simple countermeasures that could be used by any merchant to mitigate this impact.

\subsection{Analysis}
\label{sec:impact}


As shown in our experiments, the chance of successful double spending is relatively high with respect to the disconnection period.
In order to better quantify the impact of such an attack, we can now estimate what the adversary would gain from the running continuously the attack within a 10 hour period.
In this illustrative scenario, we assume that the adversary has a certain amount of currency in the system as an initial fund; to simplify the calculation we will use 1 as a number of this initial fund.
In each attempt, we let the adversary spend a third of the amount of coins available in the account; the adversary is allowed to split the amount into multiple accounts in order to transfer only one third during the attack.
If a double spending attempt is successful, the adversary will gain one third more in his account balance, which will be used as a fund for the next attempt.
To simplify the calculation further, we do not take into account any transaction processing fee. 

Overall, we can compute the potential gain after each attempt as
$ y_{i+1} = y_i(1+\frac{2p-1}{3})$, 
where $y_i$ is the adversary fund after the $i^{th}$ attack and $p$ is the attack success probability. The gain after $i$ attempts is thus 
$ y_{i} = (1+\frac{2p-1}{3})^iy_0$, where $y_0$ is the initial fund.
As shown above, the chance of attack success depends of the attack period.  Let $T$ be the duration of the attack, the potential gain  $y(t)$ at any given time $t$ then becomes:
\begin{equation}
y(t)= \left(1+\frac{2p-1}{3}\right)^{t/T}y_0.
\label{eq:gain}
\end{equation}
By exploiting the results presented in \tablename~\ref{tab:bgpsuccessrate}, we can estimate using Equation (\ref{eq:gain}) the potential gain after 10 hours of an adversary using a 9 minute attack duration as 201,903 folds of the initial balance.


\subsection{Countermeasures}
\label{sec:countermeasure}

Although it is almost impossible to completely eliminate double spending risks due to the inherent ``forkable'' design of Ethereum, we believe that there is a range of simple countermeasures that could be used to lower this risk. 

\subsubsection{Increasing the number of confirmations before considering a transaction as committed}
The simplest countermeasure already in use is for the merchant to simply wait for a higher number of confirmations, i.e., a number of subsequent blocks after the transaction appears.
Not only will the attack duration be more difficult to achieve, but it also increases the chance that the attack will be detected or disrupted by a number of configuration changes.
There is however a major drawback to this approach.
Because the merchant needs to wait for a higher number of subsequent blocks, which results in a longer period of time before sending the items, it definitely slows down the responsiveness of the system. 
%
%
While 9 minutes is sufficient to yield high profit based on the results presented in Section~\ref{sec:consortium}, once we change to 40 confirmations instead of 11 in the vanilla Ethereurm, 9 and 10 minutes attack durations with BGP hijacking result in only 3.33\% and 6.67\% success rate respectively. 
Even a 12 minutes attack duration only gives a double spending success rate  of 43.33\%. While this approach requires the attack to take longer, it cannot fully prevent it.

\subsubsection{Selectively choosing peers for query transaction status}
For an adversary to perform a double spending, the issued transaction that pays a merchant must be committed on only a group of miners but not anywhere else.
In doing so, an adversary must be aware of a group of miners that the merchant uses for a transaction verification.
For a merchant, one then needs to verify whether an issued transaction is committed before delivery of the items.
If a merchant sees that a transaction status is committed and delivers the good, there is a risk of double spending.
Before delivery, however, a merchant could mitigate the risk by simply querying a transaction status from many miners, including a group of nodes that are located further away in the network topology, such as two nodes in different ASes.
The higher number of nodes in the verification, the less chance for an adversary to alter a transaction status from the merchant's view.
Also, if a merchant cannot reach to majority of miners in the other networks or ASes, it could be a sign that there is a network partitioning. Unfortunately, the very notion of public blockchain makes this idea of majority vague and risks to limit the availability of the system.

\subsubsection{Leveraging multiple network paths}
One important requirement for the adversary to perform a man-in-the-middle attack in general is the ability to control some of the network paths.
One may mitigate such a risk by leveraging multihomed ASes and multihomed hosts for a consortium and a private blockchain environment, respectively.
Networks and hosts must be configured to allow miners to communicate with their peers using two or more network paths.
This approach makes the attack more difficult because the adversary must control more than a single network point.
While the approach does not eliminate all the risks, it requires more effort from the adversary to execute the attack.
The more network paths in used, the less likely a man-in-the-middle attack against Ethereum will succeed.

\subsubsection{Consistent blockchains}
As the environments in which  Ethereum is the most vulnerable are the consortium and private contexts, a simple approach is to use alternative blockchain systems that favor consistency over availability under the assumption that the number of malicious users is strictly lower than a third of the system. Such examples include Hyperledger Fabric~\cite{And18} that can guarantee atomic broadcast and the Red Belly Blockchain~\cite{Gra17} that solves the blockchain consensus. By favoring consistency, these blockchains first guarantee that a new block satisfies a total order over the whole set of blocks that the blockchain contains before appending it. This also means that such blockchains may not be available in the case that no such block can be found. Finally, their guarantees are not clearly defined in the case where a coalition of more than a third of participants misbehave.

\section{Related Work}
\label{sec:related}


%



Perhaps the most basic form of attack requires a transaction to be committed as soon as it is included in a block~\cite{Fin11,KAC12,BDEWW13}.
The first attack of this kind is called Finney's attack and consists of solo-mining a block with a 
transaction that sends coins to itself without broadcasting it before issuing a transaction that double-spends the same coin to a merchant. When the goods are delivered in exchange of the coins, the attacker broadcasts its block to override the payment of the merchant. 
The vector76 attack~\cite{Vec11} consists of an attacker solo-mining after block $b_0$ a new block 
$b_1$ containing a transaction to a merchant to purchase goods. 
Once another block $b_1'$ is mined after $b_0$, the attacker quickly sends $b_1$ to the merchant for
an external action to be taken. If $b_1'$ is accepted by the system, the attacker can issue another transaction with the coins spent in the discarded block $b_1$.

The attacks become harder if the external action is taken after the transaction is committed
by the blockchain. Rosenfeld's attack~\cite{Ros12} consists of issuing a transaction to a merchant. 
The attacker then starts solo-mining a longer branch while waiting for $m$ blocks to be appended 
so that the merchant takes an external action in response to the commit.  
The attack success probability depends on the number $m$ of blocks the merchant waits before taking an external action and the attacker mining power. However, when the attacker has more mining power
than the rest of the system, the attack, also called \emph{majority hashrate attack} or \emph{51-percent attack}, is guaranteed successful, regardless of the value $m$. 
To make the attack successful when the attacker owns only a quarter of the mining power, the attacker can incentivize other miners to form a coalition~\cite{EG14} until the coalition owns more than half of the total mining power.

It is well known that delaying network messages can impact Bitcoin~\cite{DW13,PSS16,SZ15,GKK16,NKMS16}.
Several attacks hence benefited from an attacker able to attack the communication 
graph~\cite{PSS16,SZ15,NKMS16}. 
Decker and Wattenhoffer already observed that Bitcoin suffered 
from block propagation delays~\cite{DW13}. 

In 2014, a BGP hijacker exploited access to an ISP to 
steal \$83000 worth of bitcoins by positioning itself between Bitcoin pools and their miners~\cite{LS14}.
At the application level, some work showed that an attacker controlling 32 IP addresses 
can ``eclipse'' a Bitcoin node with 85\% probability~\cite{HKZG15}.
Godel et al.~\cite{GKK16} analyzed the effect of propagation delays on Bitcoin using a Markov 
process. 
Garay et al.~\cite{GKL15} investigated Bitcoin in the synchronous communication setting, however, 
this setting is often considered too restrictive~\cite{Cac01}.
Pass et al. extended the analysis for when the bound-on message delivery is unknown
and showed in their model that the difficulty of Bitcoin's crypto-difficulty has to be adapted depending on the bound on the communication delays~\cite{PSS16}.

%
Even though the propagation strategy of Ethereum differs from the pull
strategy of Bitcoin, some network attacks against Bitcoin could affect
Ethereum. 
In the Eclipse attack~\cite{HKZG15} the attacker forces the victim to connect to 8 of its malicious 
identities.
The Ethereum adaptation would require to forge $3\times$ more identities and force as many 
connections as the default number of clients is 25.
Apostolaki et al.~\cite{AZV17} proposed a BGP hijacking attack and showed that
the number of Internet prefixes that need to be 
hijacked for the attack to succeed depends on the distribution of the mining power. 
BGP-hijacking typically requires the control of network operators but is independent from Bitcoin 
and could potentially be exploited to delay network messages and execute a Balance attack 
in Ethereum.


\section{Conclusion}
\label{sec:conclusion}

This paper presented the first fully implemented attack against blockchain that incorporates both components of an attack against blockchain, namely a network attack and an asset loss using double spending.
This complete attack has been deployed against the Ethereum blockchain system in different contexts. 
In the context of the public blockchain, based on real-world data, we quantified and discussed the feasibility of a network partitioning attack.
We found that while such an attack is theoretically possible, the risks of succeeding in stealing assets remains extremely low in practice. 

For the first time, we also presented an analysis of the vulnerabilities of Ethereum in both consortium and private contexts. 
When Ethereum is deployed over a WAN in a consortium environment, we demonstrated that an adversary who has control on their border-gate could easily perform a network partitioning attack, through BGP hijacking, with a double-spending success rate up-to 80\%.
Similarly, in a private environment, we showed that, using the ARP spoofing attack, an adversary could double-spend with a success rate up to 80\%.
Overall, in this two Ethereum contexts, an adversary could significantly (200,000-fold) gain by reproducing the attack continuously during 10 hours.

While the potential consequences of these attacks are impressive, we have also proposed a set of counter-measures. Short-term measures consist of increasing the number of confirmations. Long-term ones consist of monitoring network actively. Implementing active monitoring is part of our future work.

\bibliographystyle{IEEEtran}
\bibliography{reference}

\begin{thebibliography}{10}
\providecommand{\url}[1]{#1}
\csname url@samestyle\endcsname
\providecommand{\newblock}{\relax}
\providecommand{\bibinfo}[2]{#2}
\providecommand{\BIBentrySTDinterwordspacing}{\spaceskip=0pt\relax}
\providecommand{\BIBentryALTinterwordstretchfactor}{4}
\providecommand{\BIBentryALTinterwordspacing}{\spaceskip=\fontdimen2\font plus
\BIBentryALTinterwordstretchfactor\fontdimen3\font minus
  \fontdimen4\font\relax}
\providecommand{\BIBforeignlanguage}[2]{{%
\expandafter\ifx\csname l@#1\endcsname\relax
\typeout{** WARNING: IEEEtran.bst: No hyphenation pattern has been}%
\typeout{** loaded for the language `#1'. Using the pattern for}%
\typeout{** the default language instead.}%
\else
\language=\csname l@#1\endcsname
\fi
#2}}
\providecommand{\BIBdecl}{\relax}
\BIBdecl

\bibitem{Nak08}
S.~Nakamoto, ``Bitcoin: a peer-to-peer electronic cash system,'' 2008.

\bibitem{Woo15}
G.~Wood, ``Ethereum: A secure decentralised generalised transaction ledger,''
  2015, yellow paper.

\bibitem{DW13}
C.~Decker and R.~Wattenhofer, ``Information propagation in the bitcoin
  network,'' in \emph{Proc. IEEE Int. Conf. Peer-to-Peer Computing}, 2013.

\bibitem{PSS16}
R.~Pass, L.~Seeman, and A.~Shelat, ``Analysis of the blockchain protocol in
  asynchronous networks,'' Crytology ePrint Archive, Tech. Rep., 2016.

\bibitem{SZ15}
Y.~Sompolinsky and A.~Zohar, ``Secure high-rate transaction processing in
  bitcoin,'' in \emph{Proc. of {FC} 2015}, 2015, pp. 507--527.

\bibitem{GKK16}
J.~G\"obel, H.~Keeler, A.~Krzesinski, and P.~Taylor, ``Bitcoin blockchain
  dynamics: The selfish-mine strategy in the presence of propagation delay,''
  \emph{Performance Evaluation}, Juy 2016.

\bibitem{NKMS16}
K.~Nayak, S.~Kumar, A.~Miller, and E.~Shi, ``Stubborn mining: Generalizing
  selfish mining and combining with an eclipse attack,'' in \emph{Proc. of
  {IEEE} EuroS{\&}P 2016}, 2016, pp. 305--320.

\bibitem{NG16}
C.~Natoli and V.~Gramoli, ``The {Blockchain} {Anomaly},'' in \emph{2016 {IEEE}
  {NCA}}, Oct. 2016, pp. 310--317.

\bibitem{NG17}
------, ``The balance attack or why forkable blockchains are ill-suited for
  consortium,'' in \emph{IEEE/IFIP DSN'17}, Jun 2017.

\bibitem{Fin11}
\BIBentryALTinterwordspacing
H.~Finney, ``Finney's attack,'' February 2011. [Online]. Available:
  \url{https://bitcointalk.org/index.php?topic=3441.msg48384}
\BIBentrySTDinterwordspacing

\bibitem{Vec11}
\BIBentryALTinterwordspacing
vector76, ``The vector76 attack,'' August 2011. [Online]. Available:
  \url{https://bitcointalk.org/index.php?topic=36788.msg463391}
\BIBentrySTDinterwordspacing

\bibitem{HKZG15}
E.~Heilman, A.~Kendler, A.~Zohar, and S.~Goldberg, ``Eclipse attacks on
  bitcoin's peer-to-peer network,'' in \emph{{USENIX} Security}, 2015, pp.
  129--144.

\bibitem{AZV17}
M.~Apostolaki, A.~Zohar, and L.~Vanbever, ``Hijacking bitcoin: Routing attacks
  on cryptocurrencies,'' in \emph{{IEEE} {S\&P} 2017}, 2017, pp. 375--392.

\bibitem{underwood_blockchain_2016}
S.~Underwood, ``Blockchain {Beyond} {Bitcoin},'' \emph{Commun. ACM}, vol.~59,
  no.~11, pp. 15--17, Oct. 2016.

\bibitem{Ros12}
M.~Rosenfeld, ``Analysis of hashrate-based double-spending,'' 2012.

\bibitem{buterin_public_2015}
\BIBentryALTinterwordspacing
V.~Buterin, ``On {Public} and {Private} {Blockchains},'' Aug. 2015. [Online].
  Available:
  \url{https://blog.ethereum.org/2015/08/07/on-public-and-private-blockchains/}
\BIBentrySTDinterwordspacing

\bibitem{hawkinson_guidelines_1996}
J.~Hawkinson and T.~Bates, ``Guidelines for creation, selection, and
  registration of an {Autonomous} {System} ({AS}),'' Mar. 1996.

\bibitem{noauthor_db-ip_nodate}
\BIBentryALTinterwordspacing
``{DB}-{IP} - {IP} {Geolocation} and {Network} {Intelligence}.'' [Online].
  Available: \url{https://db-ip.com/}
\BIBentrySTDinterwordspacing

\bibitem{noauthor_ip_nodate-1}
\BIBentryALTinterwordspacing
``{IP} {Address} {Details} - ipinfo.io.'' [Online]. Available:
  \url{http://ipinfo.io/}
\BIBentrySTDinterwordspacing

\bibitem{noauthor_ip_nodate-2}
\BIBentryALTinterwordspacing
``{IP} {Address} {Geolocation} to trace {Country}, {Region}, {City}, {ZIP}
  {Code}, etc.'' [Online]. Available: \url{https://www.eurekapi.com/}
\BIBentrySTDinterwordspacing

\bibitem{noauthor_ip_nodate}
\BIBentryALTinterwordspacing
``{IP} {Address} to {Identify} {Geolocation} {Information}.'' [Online].
  Available: \url{http://www.ip2location.com/}
\BIBentrySTDinterwordspacing

\bibitem{noauthor_ip_nodate-3}
\BIBentryALTinterwordspacing
``{IP} {Geolocation} and {Online} {Fraud} {Prevention} {\textbar} {MaxMind}.''
  [Online]. Available: \url{https://www.maxmind.com/en/home}
\BIBentrySTDinterwordspacing

\bibitem{noauthor_caida:_nodate}
``{CAIDA}: {Center} for {Applied} {Internet} {Data} {Analysis}.''

\bibitem{noauthor_merit_nodate}
\BIBentryALTinterwordspacing
``Merit {RADb}.'' [Online]. Available: \url{http://www.radb.net/}
\BIBentrySTDinterwordspacing

\bibitem{hares_border_2006}
S.~Hares, Y.~Rekhter, and T.~Li, ``A {Border} {Gateway} {Protocol} 4
  ({BGP}-4),'' Jan. 2006.

\bibitem{heo_control_2015}
\BIBentryALTinterwordspacing
T.~Heo, ``Control {Group} v2,'' Oct. 2015. [Online]. Available:
  \url{https://www.kernel.org/doc/Documentation/cgroup-v2.txt}
\BIBentrySTDinterwordspacing

\bibitem{as_caida:_2017}
\BIBentryALTinterwordspacing
``The {CAIDA} {AS} {Relationships} {Dataset},'' Aug. 2017. [Online]. Available:
  \url{http://www.caida.org/data/as-relationships/}
\BIBentrySTDinterwordspacing

\bibitem{noauthor_quagga_nodate}
\BIBentryALTinterwordspacing
``Quagga {Software} {Routing} {Suite}.'' [Online]. Available:
  \url{http://www.nongnu.org/quagga/}
\BIBentrySTDinterwordspacing

\bibitem{ramachandran_detecting_2005}
V.~Ramachandran and S.~Nandi, ``\BIBforeignlanguage{en}{Detecting {ARP}
  {Spoofing}: {An} {Active} {Technique}},'' in
  \emph{\BIBforeignlanguage{en}{Information {Systems} {Security}}}, ser.
  Lecture {Notes} in {Computer} {Science}, 2005, pp. 239--250.

\bibitem{And18}
E.~Androulaki \emph{et~al.}, ``Hyperledger fabric: a distributed operating
  system for permissioned blockchains,'' in \emph{EuroSys}, 2018, pp.
  30:1--30:15.

\bibitem{Gra17}
\BIBentryALTinterwordspacing
V.~Gramoli, ``{T}he {R}ed {B}elly {B}lockchain,'' 2017, personal Communication,
  Facebook, USA. [Online]. Available:
  \url{http://gramoli.redbellyblockchain.io/web/doc/talks/facebook.pdf}
\BIBentrySTDinterwordspacing

\bibitem{KAC12}
G.~Karame, E.~Androulaki, and S.~Capkun, ``Two bitcoins at the price of one?
  double-spending attacks on fast payments in bitcoin,'' \emph{{IACR}
  Cryptology ePrint Archive}, vol. 2012, p. 248, 2012.

\bibitem{BDEWW13}
T.~Bamert, C.~Decker, L.~Elsen, R.~Wattenhofer, and S.~Welten, ``Have a snack,
  pay with bitcoins,'' in \emph{{IEEE} {P2P}}, 2013, pp. 1--5.

\bibitem{EG14}
I.~Eyal and E.~G. Sirer, ``Majority is not enough: Bitcoin mining is
  vulnerable,'' in \emph{Proc. of {FC} 2014}, 2014, pp. 436--454.

\bibitem{LS14}
P.~Litke and J.~Stewart, ``{BGP} hijacking for cryptocurrency profit,'' August
  2014.

\bibitem{GKL15}
J.~A. Garay, A.~Kiayias, and N.~Leonardos, ``The bitcoin backbone protocol:
  Analysis and applications,'' in \emph{34th Annu. Int. Conf. the Theory and
  Applications of Crypto. Techniques}, 2015, pp. 281--310.

\bibitem{Cac01}
C.~Cachin, ``Distributing trust on the internet,'' in \emph{Proc. Int. Conf.
  Dependable Systems and Networks ({DSN})}, 2001, pp. 183--192.

\end{thebibliography}


\appendix
\section{Demystify Ethereum's GHOST}
\label{sec:ghost}

How effective is the man-in-the-middle attack on Ethereum depends on the consensus protocol used, therefore it is essential to understand how it works under the hood.
To determine the canonical path through the entire block tree, Ethereum implements its own protocol~\cite{Woo15}.
Although stated by its author that Ethereum uses simplified version of GHOST, as purposed in ~\cite{SZ15}, we have found that there exist differences in many aspects between the two.
In this sense, knowing how such a protocol works behind the scene is important to realize the impact of any network attack.
In this section, we look deeply into the differences and also misconceptions related to Ethereum's consensus protocol.
First, we demystify its canonical branch selection process.
Then, we explain the reason how this blockchain platform takes into account of uncle blocks.
Although it sounds similar to a part of original GHOST protocol, Ethereum incorporates this part of mechanism for another purpose.
Lastly, we explain the difficulty adjustment algorithm used in the blockchain platform.

\subsubsection{Choosing a canonical branch}
Fork or branching is a common phenomenon for proof-of-work blockchains.
In order to enter new information into the system, a node needs to put the information in a form of block with a pointer to its parent block.
To create a block a node need to solve the crypto puzzle as a proof of its work; this process is called mining.
Each miner needs to adhere to the same protocol for creating and also validating new blocks.
If a miner successfully mine a block, it will broadcast to its neighbors, which can then validate the block.
Other miners may continue competing to mine more block based on the latest one onward, and a miner will be award the reward for its work.
For this incentive, all the miners will compete with each others to create new blocks to get their rewards.
Unfortunately, not all the miners will receive latest update about the global state or the latest blocks at all time; the situation leads to a fork, where there are more than one blocks point to the same parent block.

To address the fork, blockchain platforms require to implement consensus protocols.
Nakamoto consensus simply chooses the longest branch whenever there are forks, leaving blocks which are not part of the main chain as stale blocks, wasted efforts~\cite{Nak08}.
Currently, this is the protocol used in Bitcoin, but its capability to support high volume of transactions is still questionable by the community.
One of the attempts to tackle the scalability issue is a GHOST consensus protocol.
Instead of looking at the longest path, GHOST selects a canonical branch by considering total weight of the subtrees including stale blocks.

Ethereum claims that it implements a simplified version of a GHOST consensus protocol; to our knowledge, nonetheless, it is not true in several respects.
As a proof-of-work blockchain, Ethereum comes up with its own approach in order to select canonical branch.
Although Ethereum states that its implementation is based on the original GHOST protocol, it does not take into account of any stale block in the subtree when it decides the canonical path.
Instead, its branch selection process is solely based on the total difficulty value, the summation of difficulty of all the crypto puzzles of the block itself and also all of its ancestors, as a weight of each branch.
Based on this total difficulty value, Ethereum plainly picks the branch with highest accumulated weight as a canonical branch.
Unlike Nakamoto consensus or the original GHOST protocol, which treat each block equally in term of weight, a subtree with fewer number of block may be adopted as long as its total difficulty is higher than the others.

\begin{algorithm}
\caption{Total Difficulty Calculation in Ethereum}\label{algo:total_diff}
{\fontsize{9pt}{9pt}\selectfont
\begin{algorithmic}
	\Function{get-total-diff}{block}
		\If{block is genesis-block}
			\State \Return genesis-block.difficulty
		\Else
			\State \Return block.difficulty + get-total-diff(block.parent)
		\EndIf
	\EndFunction
\end{algorithmic}
}
\end{algorithm}

\subsubsection{Stale blocks and uncles}
Although it is taken into account for the weight calculation, Ethereum instead incorporates stale block rewarding system which is not a part of the original GHOST protocol.
The Ethereum's author mentions that the large mining pools tend to have an advantage over small mining pools in term of propagation delay~\cite{buterin_toward_2014}.
When any pool mine a new block, such a pool could start mining a new block right away without waiting for a block to reach other miners, thus the system tend to be centralized from this point of view.
Whenever a large mining pool successfully mine a block, it has a higher chance compare to the rest of miners to successfully mine the next block, because it could start mining ahead of them while other miners do not receive announcement of the latest block yet.
For this reason, Ethereum introduces stale block rewarding system in its consensus protocol to provide incentive to the miners who could contribute stale blocks despite those blocks are not part of canonical chain.
This mechanism helps keep the small miners stay alive.

However, Ethereum does not give reward to every stale block.
A block may optionally refer to two other stale blocks as its uncles; no more than two blocks are allowed.
Further, only stale blocks within seven generation distance from the current block can be included as uncle blocks~\cite{buterin_wiki:_2017}.
Although Ethereum gives reward to uncle blocks, those blocks are not take into account in weight calculation when selecting the head of canonical branch.
From this perspective, the higher number of uncle blocks does not help in securing or promoting higher rate of transaction.

\subsubsection{Difficulty adjustment algorithm}
Aiming to be a blockchain platform of generic applications, Ethereum's authors decide to allow its block generation at much higher rate compare to Bitcoin.
To control the rate of new blocks, both Bitcoin and Ethereum employ the similar mechanism, the difficulty of the crypto puzzle.
Each block contains the difficulty for itself to be mined as one of its attributes.
Target block time for Bitcoin is 10 minutes, whereas the most recent average block time to create a block in Ethereum is 15 seconds.
Difficulty in the former will be changed every 2016 blocks; the latter adjusts the difficulty for every new blocks.

\begin{algorithm}
\caption{Difficulty Adjustment Algorithm}\label{algo:diff_adjust}
{\fontsize{9pt}{9pt}\selectfont
\begin{algorithmic}
	\Function{get-diff-adjustment}{block, parent}
		\State fraction = $\left \lfloor{(\text{parent.difficulty} \div 2048}\right \rfloor$
		\State interval = block.timestamp - parent.timestamp
		\State factor = max(1 -  $\left \lfloor{(\text{interval} \div 10)}\right \rfloor$, -99)
		\State \Return fraction * factor
	\EndFunction
	
	\Function{get-diff-bomb}{block}
		\State exponent = $\left \lfloor{(\text{block.number} \div 100000)}\right \rfloor - 2$
		\State \Return  $\left \lfloor{2^{\text{exponent}}}\right \rfloor $
	\EndFunction
	
	\Function{get-difficulty}{block}
		\State parent = block.parent
		\State adjustment = get-diff-adjustment(block, parent)
		\State bomb = get-diff-bomb(block)
		\State \Return  parent.difficulty + adjustment + bomb
	\EndFunction
\end{algorithmic}
}
\end{algorithm}

Any block with a number greater than zero need to calculate its difficulty after solving crypto puzzle by using the value of its parent as a base.
There are two parts in difficulty update algorithm to vary the value from as the chain grow.
The first part, as illustrated in a function get-diff-adjustment in an algorithm \ref{algo:diff_adjust}, varies the difficulty based on the time a miner used to find the answer for crypto puzzle; this is calculated by subtract timestamp of its parent from timestamp of the current block.
For this part, a miner will increase the difficulty if and only if the time it takes to mine a block is less than 10 seconds.
If the amount of time falls into a range between 10 to 19, the difficulty will remain the same with its parent.
If the time a miner takes to create a block, however, greater than or equal to 20, the difficulty will be decreased.

The second part, as shown in a function get-diff-bomb, is purely based on the block number; it is known as the difficulty bomb among Ethereum community.
Because of a plan to switch from proof of work to proof of stake in the future, the developers decide to incorporate the difficulty bomb as a part of difficulty adjustment algorithm.
The role of the difficulty bomb is to exponentially increase the difficulty over time to the point that it is no longer worth for any miner to mine a block in comparison to small incentive gain.
When that point is reached, the developers will take the opportunity to switch Ethereum into proof-of-stake blockchain.

\end{document}